\documentclass[12pt]{iopart}
\usepackage{iopams}
\usepackage{amstext}
\usepackage{esint}
\usepackage{bm}
\usepackage{soul}
\begin{document}

%%%%%%%%%%%%%%%
\newcommand{\be}{\begin{equation}}
\newcommand{\ee}{\end{equation}}
\newcommand{\barr}{\begin{eqnarray}}
\newcommand{\earr}{\end{eqnarray}}

\newtheorem{thm}{Theorem}
\newcounter{tmp}
\setcounter{tmp}{\value{thm}}
\setcounter{thm}{0} 
\newtheorem{lem}{Lemma}

 \newcommand{\<}{\langle}
\renewcommand{\>}{\rangle}
\newcommand{\EE}{\mathcal{E}}
\newcommand{\R}{\mathbb{R}}
\newcommand{\de}{\mathrm{d}}
\newcommand{\XX}{\mathcal{X}}
\newcommand{\BBB}{\mathcal{B}}
\renewcommand{\rho}{\varrho}

\title[A shortcut through the Coulomb gas method]{A shortcut through the Coulomb gas method for spectral linear statistics on random matrices}

\author{Fabio Deelan Cunden$^{1}$, Paolo Facchi$^{2,3}$ and Pierpaolo Vivo$^{4}$}

\address{$1.$ School of Mathematics, University of Bristol, University Walk, Bristol BS8 1TW,  United Kingdom\\
$2.$ Dipartimento di Fisica and MECENAS, Universit\`a di Bari, I-70126 Bari, Italy\\
$3.$ Istituto Nazionale di Fisica Nucleare (INFN), Sezione di Bari, I-70126 Bari, Italy\\
$4$. King's College London, Department of Mathematics, Strand, London WC2R 2LS, United Kingdom}

\date{\today}

\begin{abstract} 
In the last decade, spectral linear statistics on large dimensional random matrices have attracted significant attention. Within the physics community, a privileged role has been played by invariant matrix ensembles for which a two dimensional Coulomb gas analogy is available. We 
present a critical revision of the Coulomb gas method in Random Matrix Theory (RMT) borrowing language and tools from Large Deviations Theory. This allows us to formalize an equivalent, but more effective and quicker route toward RMT free energy calculations. 
Moreover, we argue that this more modern viewpoint is likely to shed further light on the interesting issues of weak phase transitions and evaporation phenomena recently observed in RMT.
\end{abstract}

%Uncomment for PACS numbers title message
%\pacs{}

\maketitle
\section{Introduction}\label{sec:intro}
This article contains a critical revision of the two dimensional (2D) Coulomb gas analogy in Random Matrix Theory (RMT) and presents an alternative method to compute large deviation functions for spectral linear statistics on invariant matrix models. The Coulomb gas technique has been used in several physical problems for many decades. A more modern viewpoint on the subject, based on concepts and tools borrowed from Large Deviations Theory (LDT) seems worthwhile, as certain fundamental identities have been so far overlooked in the vast majority of previous works on this topic. 

The goal of this paper is twofold:
\begin{enumerate}
\item to present an effective shortcut to the standard Coulomb gas technique for the evaluation of probabilities of linear statistics on invariant random matrix models. The method relies on classical thermodynamic identities (\eref{eq:stip} and \eref{eq:JPsimulti} below) based on the Legendre duality~\cite{Thesis}. 

\item to argue - based on existing evidence from the huge literature on this subject - that this more modern viewpoint on an otherwise well-established technique has the potential to shed further light on phase transitions and evaporation phenomena in RMT. 
\end{enumerate}

This paper is organized as follows. Section~\ref{sec:RMT} illustrates the well-known connection between invariant matrix models and 2D Coulomb gases. For further details we refer to~\cite{Guionnet,Forrester10,Mehta04}. Section~\ref{sec:review} presents a brief review of the previous works. Section~\ref{sec:LDT} contains a few rigorous results from LDT 
that set the natural stage for our discussion. The reader interested in the details of the proofs is advised to look at~\cite{Ellis99}. 
The large-$N$ limit of matrix integrals is discussed in Section~\ref{sec:largeN}  and the general ideas of the Coulomb gas method are presented in Section~\ref{sec:method} through the prism of LDT.
Section~\ref{sec:shortcut} contains the thermodynamic identities for linear spectral statistics and their use to compute large deviation functions; this section also contains simple, convincing examples illustrating the power of the shortened method and a brief discussion on the thermodynamical meaning of the Legendre duality. In addition to the shortcut, our investigations naturally prompt a corpus of new ideas on the issues of phase transitions and evaporation phenomena in Coulomb gas systems. These ideas and related open problems are discussed at length in Section~\ref{sec:discussion}.

\section{Random matrices and 2D Coulomb gases}\label{sec:RMT}
The definition of a matrix ensemble consists of:
\begin{itemize}
\item[i)] The choice of a set of matrices $\mathcal{M}_{\beta}$. 
We shall consider the following sets:  
real $N\times N$ symmetric matrices; 
complex $N\times N$ Hermitian matrices; 
quaternion $N\times N$ self-dual matrices. 
They are conveniently labeled by the Dyson index $\beta=1,2$ and $4$,  respectively.
\item[ii)] The choice of a probability measure on $\mathcal{M}_{\beta}$.
Each of the sets with $\beta=1,2,4$ 
is naturally equipped with its own Lebesgue measure $\mu_L$. However, for these sets $\mu_L$ is infinite and hence has no probabilistic meaning. Instead, we shall consider the probability measure
\be
\de\mu(M)=K_{N,\beta}\exp\left[-\beta N \Tr V(M)\right]\de\mu_L(M).
\label{eq:measure}
\ee
$M$ denotes a generic matrix in $\mathcal{M}_{\beta}$, 
$V$ is a real-valued function on $\mathcal{M}_{\beta}$ 
such that the measure is normalizable and  $K_{N,\beta}$ is a normalization constant. This measure is invariant under the adjoint action of the group $\mathcal{U}_{\beta}$, 
where  
$\mathcal{U}_{\beta}=O(N), U(N)$ or $Sp(N)$ for $\beta=1,2$ or $4$ respectively. In formulae, for any $U\in\mathcal{U}_{\beta}$ 
\be
\de\mu(M)=\de\mu(U^{-1}MU). \label{eq:invariance}
\ee
\end{itemize}
These ensembles are usually called \emph{invariant ensembles}.
For the physical motivations behind the choices of $\mathcal{M}_{\beta}$ 
as in i) we refer to the second Chapter of Mehta's book~\cite{Mehta04}. Matrices with $\beta=1,2,4$ 
are diagonalizable with real spectrum. As a consequence of the `rotational' invariance~\eref{eq:invariance}, eigenvalues and eigenvectors of $M$ are statistically independent, and any basis is `equally likely' to diagonalize $M$. Therefore, the eigenvectors of invariant ensembles are quite immaterial. On the other hand, with the choice~\eref{eq:measure} as probability measure on $\mathcal{M}_{\beta}$ 
the joint law of the eigenvalues $(\lambda_1,\dots, \lambda_N)$ is~\cite{Mehta04} 
\be
\de P(\lambda_1,\dots,\lambda_N)=Z_{N,\beta}^{-1}\,e^{-\beta E(\lambda_1,\dots,\lambda_N)}\de\lambda_1\cdots \de\lambda_N ,
\label{eq:jpdf}
\ee
with
\be
 E(\lambda_1,\dots,\lambda_N)=-\frac{1}{2}\sum_{i\neq j}\log|\lambda_i-\lambda_j|+N\sum_{i}V(\lambda_i), 
 \label{eq:energy}
\ee
where $Z_{N,\beta}=\int d\lambda_1\cdots d\lambda_N e^{-\beta E(\lambda_1,\dots,\lambda_N)}$ is a normalization constant and hereafter, unless otherwise specified, all summations run from $1$ to $N$.  It is evident that~\eref{eq:jpdf} is the canonical distribution of a system of identical particles on the line labeled by their positions $\lambda_1,\dots,\lambda_N$ at inverse temperature $\beta$. These $N$ particles interact (logarithmic repulsion) and experience a confinement by the external potential $NV(\lambda)$. Note that the logarithmic interaction is solution of the 2D Poisson equation. For this reason, the system is referred to as a 2D Coulomb gas constrained on the line (the eigenvalues are real) and the normalization constant $Z_{N,\beta}$ acquires  the meaning of a (positional) partition function. 

This Coulomb gas picture is a useful and physically intuitive device for making educated guesses on the behavior of invariant ensembles. In view of certain applications to physics (see, e.g.,~\cite{Forrester10,Serfaty14}), it is also worth studying the canonical measure~\eref{eq:jpdf}-\eref{eq:energy} of a particle system with logarithmic pair repulsion at inverse temperature $\beta>0$ independently of a concrete matrix model realization. Note that explicit (non invariant) matrix models with arbitrary non-quantized index $\beta>0$ are known in RMT~\cite{Dumitriu02}.

Let us first start with a quick review of the method and its wide range of applications so far.

\section{Review of the Coulomb gas method}\label{sec:review}
The basic idea of the Coulomb gas picture goes back to the works of Wigner~\cite{Wigner57} in the 1950s. However, it was with the formidable series of papers by Dyson~\cite{Dyson62} in the 1960s that the exact correspondence between the eigenvalue distributions of some random matrix models and the statistical mechanics of classical 2D Coulomb gases attracted the attention of physicists. At the heart of the  Coulomb gas method lies the following observation~\cite{Dyson62}: the probability law of a spectral statistics $F=f(\lambda_1,\dots,\lambda_N)$ on an invariant matrix ensemble can be expressed as a ratio of two partition functions
\barr
\Pr(F\in\mathcal{B})=\int_{\mathcal{B}}\de a\int\cdots\int\delta[a-f(\lambda_1,\dots,\lambda_N)] \de P(\lambda_1,\dots,\lambda_N)\\
=\frac{1}{Z_{N,\beta}}\int  \limits_{ f(\lambda_1,\dots,\lambda_N)\in\mathcal{B}}  \hspace{-7mm} \de\lambda_1\cdots\int \de\lambda_N\; e^{-\beta E(\lambda_1,\dots,\lambda_N)}=\frac{Z_{N,\beta}(\mathcal{B})}{Z_{N,\beta}}\ .
\label{eq:ratio}
\earr
The numerator in~\eref{eq:ratio} is the partition function of the associated Coulomb gas with the constraint $F\in\cal{B}$; the denominator is the partition function of the unconstrained Coulomb gas.

Dyson made the assumption that \emph{for large $N$ the Coulomb gas obeys the laws of classical thermodynamics}. He obtained many results based more on well-rooted statistical mechanics intuitions rather than arguments of a mathematically rigorous kind. Later, several aspects of the problem have been put on a more formal ground thanks to the development of modern ideas of LDT. However, this program is far from being completed, as we will try to argue below.

The seminal idea of Dyson has been rediscovered several times in the past and pushed forward to compute the large deviation functions of spectral statistics on matrix models. The first remarkable ``Coulomb gas calculations'' appear in the works~\cite{Brezin78,Gross80,Douglas93} on the large-$N$ approximation in various gauge theories. More recently, a version of the technique with hard-wall constraints has been engineered to compute large deviations of the extreme eigenvalues of Gaussian matrices~\cite{Dean06}. Scientists from several areas have recognized a 2D Coulomb gas picture in several unrelated problems such as spectral statistics 
in classical matrix models~\cite{Chen94,Fogler95,Vivo07,Majumdar09,Majumdar11,Vivo12,Majumdar13,Marino14,Cunden14}; distribution of entanglement measures in bipartite quantum systems~\cite{Page93,Facchi08,DePasquale10,Nadal10,Facchi13}; distribution of linear statistics on the transmission matrix and time-delay matrix of mesoscopic cavities~\cite{Vivo08,Texier13,Grabsch15,Cunden15}; distribution of maximal displacements for vicious random walkers in one dimension~\cite{Nadal09}; non-local correlation functions in domino tilings~\cite{Colomo13}; thermodynamics of the six-vertex model~\cite{Colomo15}; fluctuations in the 2D one-component plasma~\cite{Allez14,Cunden15b}, and many others.

However, a revision of the existing literature reveals that the full arsenal of statistical mechanics tools (some of which were at the heart of Dyson's calculations) has been severely underemployed thus far (with the exception of~\cite{Cunden14,Grabsch15,Cunden15,Cunden15b}). Since many of these stat-mech considerations have been lately formalized as LDT theorems, it seems appropriate to first summarize the relevant rigorous results. Next, we will use them to offer a modern take on various statements scattered in the physics literature.

\section{Notions of large deviation theory}\label{sec:LDT}
Large deviations theory is an asymptotic theory which deals with probability of `rare events'. See~\cite{Ellis99,Guionnet04,Touchette09}. 
Let $\XX$ be a  \emph{Polish space}, i.e.\ a complete separable metric space.  A sequence of probability measures $\Pr_N$ on  $\XX$ satisfies a \emph{large deviation principle} (LDP for short) with \emph{speed} $v_N$ and \emph{rate function} $I$ if:  i) $v_N\to\infty$ (as $N\to\infty$); ii) $I\colon\mathcal{X}\to[0,\infty]$ is lower semicontinuous; and iii) for all Borel measurable sets $\BBB$ of $\mathcal{X}$
\begin{equation}
-I(\mathring{\BBB})\leq\liminf_{N\to\infty}\frac{1}{v_N}\log{\Pr}_{N}(\BBB)\leq\limsup_{N\to\infty}\frac{1}{v_N}\log{\Pr}_{N}(\BBB)\leq-I(\bar{\BBB}) .
\label{eq:def_LDP}
\end{equation}
Here, $\mathring{\BBB}$ and $\bar{\BBB}$ are the interior and the closure of $\BBB$, respectively, and we used the notation $I(S)=\inf_{x\in S}I(x)$.
In the following we will use the shorter notation 
\begin{equation}
{\Pr}_N(\BBB)\approx \exp(-v_N I(\BBB)), \qquad N\to\infty,
\end{equation} 
for expressing the chain of inequalities~\eref{eq:def_LDP}.
For a sequence of random variables $X_N$ on $\XX$ we will loosely speak of `LDP of $X_N$', meaning the LDP for the sequence of probability laws of $X_N$. 

A simple example of Polish space is the familiar $\mathcal{X}=\R^n$. A class of Polish spaces arising naturally in applications is obtained by taking a Polish space $\mathcal{X}$ and then considering the set of probability measures $\mathcal{M}(\mathcal{X})$ on it (equipped with the L\'evy metric). In fact, in what follows we shall concern with the cases $\XX=\R^n$ and $\XX=\mathcal{M}(\R^n)$.

We recall here two practical ideas which are commonly used to prove a LDP. The first one is quite general: a LDP  can be inferred from another one.
\begin{lem}[Contraction principle]\label{lem:contraction} Let $\XX$ and $\mathcal{Y}$ be Polish spaces and $g\colon\mathcal{X}\to\mathcal{Y}$ be a measurable map. If the $\mathcal{X}$-valued random variables $X_N$ satisfies a LDP with speed $v_N$ and rate function $I$ and $g$ is continuous on $\{x\in\mathcal{X}\colon I(x)<\infty\}$, then $Y_N=g(X_N)$ satisfy a LDP with same speed and rate function 
\be
\Psi(y)=\inf\{I(x)\colon\; g(x)=y\}.
\ee
\end{lem}
Notice that the above lemma is nothing but an instance of the so-called saddle-point approximation. 

A somehow converse statement would be the following. Suppose 
that for a large class of functions $F\colon\XX\to\R$ the limit   
\be
\lim_{N\to\infty}(1/v_N)\log \<\exp(v_N F(X_N))\> \label{eq:Bryc}
\ee 
exists ($\<{\cdot}\>$ denotes the expectation with respect to the probability law of the random variable $X_N$). Then we might expect that $X_N$ satisfies a LDP with speed $v_N$.
In fact an exhaustive class of functions $F$ is known~\cite{Bryc90}, but for practical purposes that is too large. Amazingly, in the finite dimensional case $\XX=\R^n$, G\"artner~\cite{Gartner77} and Ellis~\cite{Ellis84} have shown that it is sufficient to check~\eref{eq:Bryc} for linear functions only (provided some regularity is assumed). And in this case it is also possible to reconstruct the rate function.
\begin{lem}[G\"artner-Ellis theorem for random vectors]\label{lem:GE} Let $\vec{X}_N$ be a real random vector in $\R^n$. Let $\mathcal{D}$ be the set of $\vec{s}\in\R^n$ such that the following limit exists
	\be
	J(\vec{s})=-\lim_{N\to\infty}\frac{1}{v_N}\log\< \exp({-v_N\vec{s}\cdot\vec{X}_N})\>.
	\label{eq:Jdef}
	\ee
[Note that $\vec{0}\in\mathcal{D}$ (hence $\mathcal{D}\neq\emptyset$) and $J(\vec{s})$ is concave (by H\"older inequality)]. If i) $\vec{0}\in\mathring{\mathcal{D}}$, ii) $J(\vec{s})$ is differentiable in $\mathring{\mathcal{D}}$ and iii) $\|\vec{\nabla}J(\vec{s})\|\to\infty$ on the boundary points of $\mathcal{D}$, then $\vec{X}_N$ satisfies a LDP with speed $v_N$. The rate function $\Psi(\vec{x})$ is the Legendre-Fenchel transform of $J(\vec{s})$:
	\be
  \Psi(\vec{x})=\sup_{\vec{s}\in\mathcal{D}}[J(\vec{s})-\vec{s}\cdot\vec{x}]\ .
  \label{eq:LFdef}
	\ee
\end{lem}
	The function $J(\vec{s})$ is known as \emph{scaled cumulant generating function} (cumulant GF for short) of $\vec{X}_N$. Both the cumulant generating function $J(\vec{s})$ and the rate function $\Psi(\vec{x})$ are generically called \emph{large deviation functions}.

The rate function $\Psi(\vec{x})$ provides information not only about the large deviations, but also about typical fluctuations and the most probable values which
correspond to the zeros (global minima) of $\Psi(\vec{x})$. In the case where $\Psi(\vec{x})$ has only
one global minimum, the most probable value is also the typical value, that is $\vec{X}_N$ satisfy a Law of Large Numbers for large $N$.
Moreover, if $J(\vec{s})$ is analytic at $\vec{s}=\vec{0}$, the mixed \emph{cumulant} of  $\vec{X}_N=(X_{N,1},\dots,X_{N,n})$ at leading order in $N$ are given by  
\be
\<X_{N,1}^{m_1}\cdots X_{N,n}^{m_n}\>_c=\left(-\frac{1}{v_N}\right)^{m_1+\cdots+m_n-1}\!\!\partial_{ s_1}^{m_1}\cdots \partial_{s_n}^{m_n}J(\vec{s})\Bigl|_{\vec{s}=\vec{0}},
\label{eq:mixed_cumulants0}
\ee
where $\<\cdot\>_c$ denotes the connected average (this is in fact the origin of the name `cumulant GF'). 

The requirement \emph{iii)} in G\"artner-Ellis theorem is known as \emph{steepness condition}, and it has to be checked whenever $\mathcal{D}\neq\R^n$ (if $J(\vec{s})$ exists for all $\vec{s}\in\R^n$, then condition \emph{iii)} is trivially true). Although the steepness condition might appear a rather technical hypothesis, it is in fact a very natural requirement valid, for instance, if $J(\vec{s})$ is everywhere differentiable in its domain $\mathcal{D}$ (not only in $\mathring{\mathcal{D}})$. It may however happen that $\|\vec{\nabla}J(\vec{s})\|$ does not diverge as $\vec{s}$ approaches the boundary of $\mathcal{D}$. In these cases, only a `local version' of the G\"artner-Ellis theorem holds and, in general, the Legendre-Fenchel transform of $J(\vec{s})$ provides only the `convex envelope' of $\Psi(\vec{x})$.  One should always check whether$J(\vec{s})$ is steep or not, as the failure of this condition might have deep consequences on the statistics of $\vec{X}$ as we will discuss later for the case of spectral statistics.  

As a final remark, we remind that the Legendre-Fenchel transform~(\ref{eq:LFdef}) yields a convex function. A nontrivial fact is that a rate function $\Psi(\vec{x})$ obtained from the G\"artner-Ellis theorem is necessarily \emph{strictly convex}~\cite{Rockafellar}.

We are now ready to briefly revisit the standard route followed in the physics literature thus far in the light of the rigorous results above. This should lead quite naturally to the proposed shortened route.

\section{Limit of large size matrices}\label{sec:largeN}
One of the central problem in RMT is to find methods to compute averages with respect to the probability measure~(\ref{eq:measure}) on 
the invariant ensemble $\mathcal{M}_{\beta}$ 
\be
\int F(M)\de\mu(M)=K_{N,\beta} \int F(M)e^{-\beta N\Tr V(M)}\de\mu_L(M).
\ee
Suppose that $F(M)=f(\lambda_1,\dots,\lambda_N)$ depends only on the spectrum $(\lambda_1,\dots,\lambda_N)$ of $M$. The `rotational' invariance~\eref{eq:invariance} of the measure allows us to integrate out the angular degrees of freedom and get
\be
\int F(M)d\mu(M)=Z_{N,\beta}^{-1} \int f(\lambda_1,\dots,\lambda_N)e^{-\beta E(\lambda_1,\dots,\lambda_N)}\de\lambda_1\cdots \de\lambda_N. \label{eq:integral}
\ee
In the Coulomb gas picture, this corresponds to computing averages of thermodynamic quantities. 
In the majority of cases, such averages are hard to compute for finite $N$. It is quite natural, therefore, to look for simpler and more convenient evaluations of these averages for large $N$. 

Clearly, at zero temperature $\beta\to\infty$ the only contribution to the matrix integral~\eref{eq:integral} comes from the configurations of the gas which minimize the energy function $E(\lambda_1,\dots,\lambda_N)$. At finite temperature $\beta<\infty$, one has to look for asymptotic formulae valid when the number of particles of the Coulomb gas is large. In the thermodynamic limit, the task of computing \emph{averages} is replaced by \emph{most probable values} assumed to be approximately equal to the corresponding averages. In the large $N$ limit, a concentration of measure phenomenon indeed makes the thermodynamic variables close to their averages: the most probable value is also the \emph{typical value}.

These ``classical" arguments can now be stated rigorously as a LDP for the configurations of the Coulomb gas. 
The eigenvalues $(\lambda_1,\dots,\lambda_N)$ (microstates of the system) are conveniently described in terms of the normalized empirical measure (macrostates)
\be
\rho_N(\lambda)=N^{-1}\sum_{i}\delta(\lambda-\lambda_i).
\label{eq:empmeas}
\ee
For large $N$ the energy function has a mean-field limit 
\begin{equation}
E(\lambda_1,\dots,\lambda_N)= N^2 \EE[\rho_N]+o(N^2), \qquad N\to\infty
\end{equation} 
where the mean-field energy density functional is  
\be
\EE[\rho]=-\frac{1}{2}\iint\limits_{\lambda\neq \lambda'}\log|\lambda-\lambda'| \de\rho(\lambda)\de\rho(\lambda')+\int V(\lambda)\de\rho(\lambda).\label{eq:functional}
\ee
This functional enjoys many nice analytical properties. Most importantly, under some mild assumptions on $V(\lambda)$, there exists a unique probability measure $\rho^*$ that minimizes $\EE[\rho]$, namely
\be
\EE[\rho^*]=\inf\left\{\EE[\rho]\,\colon\;\rho\geq0,\,\, \int \de\rho(x)=1\right\}.
\ee
In this case the following result due to Ben Arous and Guionnet~\cite{Guionnet97,Guionnet04,Guionnet} holds (see also~\cite{Chafai}).
\begin{thm}[LDP for the empirical measure]\label{thm:LDP} If $\liminf_{|\lambda|\to\infty}V(\lambda)/\log|\lambda|>1$, then $\rho_N$ satisfies a large deviation principle with speed $\beta N^2$ and rate function 
\be
\Delta\EE[\rho]=\EE[\rho]-\EE[\rho^*].
\label{eq:DEdef}
\ee
\end{thm}
A physical translation of Theorem~\ref{thm:LDP} is as follows. If  in~(\ref{eq:energy}) the external potential $V(\lambda)$ grows faster than the logarithmic repulsion (so that a confinement of the Coulomb gas really occurs), then the probability of a configuration $\rho_N$ is $\Pr[\rho_N]\approx \exp({-\beta N^2 \Delta\EE[\rho_N]})$. This probability is exponentially
suppressed with speed $\beta N^2$ and the precise measure of ``unlikeliness" of an out-of-equilibrium configuration is given by the energy penalty $\Delta\EE[\rho]$ with respect to the equilibrium configuration in the thermodynamic limit $\rho^*$. It is remarkable that for a system of $N$ particles, the rate for this suppression is $\Or(N^2)$. 

Hence, from the LDP, the matrix integral~\eref{eq:integral} can be estimated as the value of $f$ computed at the minimizers of the energy function $E(\lambda_1,\dots,\lambda_N)$. The accuracy of this saddle-point approximation increases exponentially in $N^2$.
In fact, once a LDP for the empirical measure of the gas is established, using the contraction principle (Lemma~\ref{lem:contraction}) one can deduce a LPD for observables of the gas, i.e. spectral statistics. 

In the following, we will be mainly concerned with \emph{linear spectral statistics} (or linear statistics for short), i.e.\ functions on the spectrum of $M$ of the form
\be
F(M)=N^{-1}\Tr f(M)=N^{-1}\sum_{i}f(\lambda_i)
\label{eq:F(M)def}
\ee
for some continuous real-valued $f(x)$. We denote by $\mathcal{P}_N(x)$ the probability distribution of $F(M)$. Note that a linear spectral statistics is a linear functional of the empirical measure~(\ref{eq:empmeas})
\be
F(M)\equiv F[\rho_N]=\int f(\lambda)\de\rho_N(\lambda)\ .
\ee 
From Lemma~\ref{lem:contraction}, if the functional $F\mathopen{[}\rho\mathclose{]}$ is continuous on the set $\{\rho\,\colon \Delta\EE[\rho]<\infty\}$, then $F$ satisfies a LDP $\mathcal{P}_N(x)\approx\exp(-v_N\Psi(x))$ with speed $v_N=\beta N^2$ and rate function
\be
\Psi(x)=\inf\{\Delta\EE\mathopen{[}\rho \mathclose{]}\,\colon\; F\mathopen{[}\rho\mathclose{]}=x\}.
\label{eq:ratefunction}
\ee
The generalization to more linear statistics is immediate. For $n>1$ linear statistics \begin{equation}
\vec{F}(M)=(F_1(M),\dots, F_n(M))=N^{-1}(\Tr f_1(M),\dots,\Tr f_n(M))
\label{eq:nlinstat}
\end{equation} 
continuous on the probability measures $\rho$ with finite mean-field energy, a LDP with speed $v_N=\beta N^2$ and \emph{joint} rate function
\be
\Psi(\vec{x})=\inf\{\Delta\EE\mathopen{[}\rho \mathclose{]}\,\colon\; \vec{F}\mathopen{[}\rho\mathclose{]} = \vec{x}
\}
\ee
holds. How to compute these rate functions? 
\section{Coulomb gas method and linear spectral statistics}\label{sec:method}
The strategy to compute the rate function of linear statistics is the following. For definiteness we shall discuss the case of a \emph{single} linear statistics $F(M)$. The idea is to compute first its cumulant GF, as in~(\ref{eq:Jdef}),
\be 
J(s)= 
-\lim_{N\to\infty}\frac{1}{\beta N^2}\log\int e^{-\beta N^2 s F(M)}\de\mu(M)
\ee 
and then recover the rate function $\Psi(x)$ as Legendre-Fenchel transform by the G\"artner-Ellis theorem (Lemma~\ref{lem:GE}).  As in Section~\ref{sec:review}, the Laplace transform $\widehat{\mathcal{P}}(s)$ of the probability law of $F(M)$ can be cast as a ratio of two partition functions
\barr
\widehat{\mathcal{P}}(s)&=\int e^{-\beta N^2 s F(M)}\de\mu(M)\\
&=\frac{\int\cdots\int e^{-\beta [E(\lambda_1,\dots,\lambda_N)+s N\sum_{k}f(\lambda_k)]}\de\lambda_1\cdots \de\lambda_N}{\int\cdots\int e^{-\beta E(\lambda_1,\dots,\lambda_N)}\de\lambda_1\cdots \de\lambda_N}=\frac{Z_{N,\beta}(s)}{Z_{N,\beta}(0)} ,
\earr
where we have suitably rescaled the Laplace variable for later convenience. Here
\be
Z_{N,\beta}(s)=\int\cdots\int \exp[{-\beta E(\lambda_1,\dots,\lambda_N;s)}]\de\lambda_1\cdots \de\lambda_N
\label{eq:zns}
\ee 
is the partition function of the original Coulomb gas subject to an additional single-particle potential $s N f(\lambda)$:
\barr
 E(\lambda_1,\dots,\lambda_N;s)&=E(\lambda_1,\dots,\lambda_N)+s N \sum_i f(\lambda_i)\\
&=-\frac{1}{2}\sum_{i\neq j}\log|\lambda_i-\lambda_j|+N\sum_{i}[V(\lambda_i)+sf(\lambda_i)].
\earr
Note that $Z_{N,\beta}(0)$ is the partition function in the absence of the additional potential, and therefore coincides with the normalization constant $Z_{N,\beta}$ in~\eref{eq:jpdf}. 
Thus the cumulant GF $J(s)$ may be expressed as \emph{excess free energy} of the 2D constrained Coulomb gas with respect to the unperturbed system
\be
\fl\qquad J(s)=-\lim_{N\to\infty}\frac{1}{\beta N^2}\log\widehat{\mathcal{P}}(s)=- \lim_{N\to\infty}\frac{1}{\beta N^2}[\log Z_{N,\beta}(s)-\log Z_{N,\beta}(0)]. \label{eq:excess}
\ee
The existence of the $N\to\infty$ limit above is ensured by the LDP in Theorem~\ref{thm:LDP} and Lemma~\ref{lem:contraction}. Hence, the computation of the joint cumulant GF $J(s)$ is tantamount to evaluating the leading order in $N$ of the partition function $Z_{N,\beta}(s)$. 
The LDP of the empirical measure $\rho_N$ states that for large $N$ the partition function $Z_{N,\beta}(s)$ in~(\ref{eq:zns}) is  
 dominated  by $\rho^*_s(\lambda)$, the unique minimizer of an \emph{effective} mean-field energy functional
$\mathcal{E}_s[\rho]$ in the space of normalized distributions:
\barr
\fl\quad  Z_{N,\beta}(s)\approx \exp\left({-\beta N^2 \mathcal{E}_s[\rho^*_s]}\right),\label{eq:largeZ}\quad
\text{with}\quad\mathcal{E}_s[\rho^*_s]=\inf\Big\{\mathcal{E}_s[\rho] \,:\, \rho\geq0 , \int \de\rho(x)=1\Big\}.
\label{eq:LDPzns}
\earr
The effective energy functional  
is 
\barr
\EE_s[\rho]&=&\mathcal{E}[\rho]+sF[\rho]
\label{eq:EE_s}\\
&=&-\frac{1}{2}\iint\limits_{\lambda\neq \lambda'}\log|\lambda-\lambda'| \de\rho(\lambda)\de\rho(\lambda')+\int [V(\lambda)+s f(\lambda)]\de\rho(\lambda)\label{eq:functional_eff}
\earr
and $\mathcal{E}_0[\rho]=\mathcal{E}[\rho]$ is given in~(\ref{eq:functional}). By definition, the saddle-points of $\EE_s[\rho]$ obey $\delta \EE_s[\rho^*_s]/\delta\rho^*_s=0$, that is 
\be
\int\log|\lambda-\lambda'|\de\rho^*_s(\lambda')-[V(\lambda)+sf(\lambda)]=C\label{eq:optim}
\ee
for some constant $C$, and for all $\lambda$ belonging to the support of $\rho^*_s$, namely $\lambda\in\mathrm{supp}\rho^*_s$. At equilibrium, the 2D Coulomb gas arranges itself in such a way that each particle  has equal electrostatic energy (the left hand side of~\eref{eq:optim}). Otherwise stated, there is no net electric field among the charges  (a necessary condition for electrostatic equilibrium). There exist many techniques to solve~\eref{eq:optim} and we shall not discuss them here, referring instead to~\cite{Tricomi57,Forrester10}. Notice that~\eref{eq:optim} is \emph{not} a \emph{linear} integral equation. Indeed, the  unknown equilibrium measure $\rho^*_s(\lambda)$ must satisfy the identity~\eref{eq:optim} for $\lambda$ in its support $\mathrm{supp}\rho^*_s$, which is itself unknown.

 The meaning of the equilibrium density is the following: $\rho^*_{s}(\lambda)$ corresponds to the \emph{typical} configuration of eigenvalues yielding 
a prescribed value of $F(M)$ as given by
\be
F[\rho^*_s]=\int f(\lambda)\de\rho^*_s(\lambda) = x^*(s)\ .
\label{eq:f(s)} 
\ee
In the language of statistical mechanics, $\rho^*_{s}(\lambda)$ is the macrostate of the system that realizes the (unlikely) event $F(M)=x$ in the most likely way, where $x=\int f(\lambda)\de\rho^*_s(\lambda)$. 
Hence, a possible route to evaluate $J(s)$ is to find the saddle-point density $\rho^*_{s}(\lambda)$ as a function of $s$ (as solution of~\eref{eq:optim}) and to insert it back into the energy functional~\eref{eq:functional_eff} to evaluate the leading order of $Z_{N,\beta}(s)$ as excess free energy in~\eref{eq:excess}. Using~\eref{eq:largeZ} we have
\be
J(s)=\EE_s[\rho^*_s]-\EE[\rho^*].\label{eq:Jdiff}
\ee
 Once $J(s)$ is known, the rate function $\Psi(x)$  of the LDP $\mathcal{P}_N(x)\approx\exp(-\beta N^2\Psi(x))$ can be computed as Legendre-Fenchel transform~(\ref{eq:LFdef}) of $J(s)$. This technique has been exploited in the last decade in many physical problems to compute the large deviations of single observables. However, this route entails the explicit computation of the mean-field energy at the saddle-point density~$\EE_s[\rho^*_s]=-\frac{1}{2}\iint\limits\log|\lambda-\lambda'| \de\rho_s^*(\lambda)\de\rho_s^*(\lambda')+\int [V(\lambda)+s f(\lambda)]\de\rho_s^*(\lambda)$, which is not necessarily an easy task. The situation gets even worse in the case of joint statistics. For $n>1$ linear statistics~(\ref{eq:nlinstat})
one has
\be
J(\vec{s})=\EE_{\vec{s}}[\rho^*_{\vec{s}}]-\EE[\rho^*]
\ee
where $\vec{s}\in\R^n$ and $\rho^*_{\vec{s}}(\lambda)$ is the minimizer of the effective energy functional
\be
\EE_{\vec{s}}[\rho]=-\frac{1}{2}\iint\limits_{\lambda\neq \lambda'}\log|\lambda-\lambda'| \de\rho(\lambda)\de\rho(\lambda')+\int [V(\lambda)+\vec{s}\cdot\vec{f}(\lambda)]\de\rho(\lambda),\label{eq:functional_eff_multi}
\ee
with $\vec{f}(\lambda)=(f_1(\lambda),\dots,f_n(\lambda))$.

Can we bypass at all the explicit and often cumbersome evaluation of the integrals in $\EE_s[\rho^*_s]$ or $\EE_{\vec{s}}[\rho^*_{\vec{s}}]$?

\section{The Coulomb gas method via Legendre duality}\label{sec:shortcut}
It is indeed possible to introduce a shortcut (based on classical thermodynamic identities, discussed extensively in~\cite{Thesis}) which outmaneuvers unwieldy integrals altogether.  We discuss first the case  of one linear statistics $n=1$. The extension to $n>1$ linear statistics as well as a discussion on the thermodynamic meaning of certain identities is then presented.

\subsection{Shortcut}
If $J(s)$ satisfies the hypotheses of the G\"artner-Ellis theorem (Lemma~\ref{lem:GE}), then the rate function of the LDP of the linear statistics $F(M)$ is given by
\barr
\Psi(x)=\sup_{s}\left[J(s)-sx\right]&=\sup_{s}\left\{\EE_{s}\mathopen{[}\rho^{*}_{s}\mathclose{]}-\EE\mathopen{[}\rho^{*}\mathclose{]}-sx\right\}\label{eq:Legendrea}\\
&=\sup_{s}\Big\{\inf_{\rho}\EE_{s}\mathopen{[}\rho\mathclose{]}-\EE\mathopen{[}\rho^{*}\mathclose{]}-sx\Big\}\\
&=\sup_{s}\Big\{\inf_{\rho}\left(\EE\mathopen{[}\rho\mathclose{]}+s F\mathopen{[}\rho\mathclose{]}\right)-\EE\mathopen{[}\rho^{*}\mathclose{]}-sx\Big\}\\
&=\sup_{s}\inf_{\rho}\{\Delta\EE\mathopen{[}\rho\mathclose{]}-s\left(x-F\mathopen{[}\rho\mathclose{]}\right)\}\ ,
\label{eq:Legendre}
\earr
where we applied (in  order)~(\ref{eq:LFdef}), (\ref{eq:Jdiff}), (\ref{eq:LDPzns}), (\ref{eq:EE_s}), and~(\ref{eq:DEdef}). 
On the other hand, from the LDP of the empirical measure of the Coulomb gas (Theorem~\ref{thm:LDP}) and the contraction principle (Lemma~\ref{lem:contraction}), the rate function is given by~(\ref{eq:ratefunction}):
\be
\Psi(x)=\inf_{\rho}\left\{\Delta\EE\mathopen{[}\rho\mathclose{]}\,:\,F\mathopen{[}\rho\mathclose{]}=x\right\}\ ,\label{eq:rate_inf_mu}
\ee
where the infimum is taken over the measures with a prescribed value 
\begin{equation}
F[\rho]=\int{f(\lambda)\de\rho(\lambda)}=x\ .
\end{equation} 
We see that the formulation in the Laplace space \eref{eq:Legendre}, is nothing but the implementation of the constrained minimization problem \eref{eq:rate_inf_mu} with a Lagrange multiplier $s$ (the Laplace variable) that takes into account the constraint $F[\rho]=x$. 

We have already remarked that $J(s)$ is concave. This implies that $J(s)$ is continuous in $\mathring{\mathcal{D}}$, and is differentiable almost
everywhere. In the following we assume further that $J(s)$ is \emph{strictly concave} and \emph{everywhere differentiable}. Since $\Psi(x)$ in~\eref{eq:Legendrea} is always strictly convex we can apply again the Legendre-Fenchel transform to get
\be
J(s)= \inf_x[\Psi(x)+sx]. \label{eq:JPsi}
\ee
In the case of regular functions, the Legendre-Fenchel transform~(\ref{eq:JPsi}) is given by 
\begin{equation}
J(s)=\Psi(x^*(s))+sx^*(s),
\label{eq:JPsi1}
\end{equation} 
where $x^*(s)$ is the unique solution of
$\Psi^{\prime}(x) = -s$.
Analogously, the transform~(\ref{eq:Legendrea}) reads
\begin{equation}
\Psi(x)=J(s^*(x))- s^*(x) x,
\label{eq:PsiJ1}
\end{equation} 
where $s^*(x)$ is the  solution of
$J^{\prime}(s) = x$.
Therefore, we get 
\begin{equation}
x^{*}(s) = J^\prime(s), \qquad s^{*}(x)=-\Psi^{\prime}(x), 
\label{eq:slopes}
\end{equation}
and the function $s^{*}(x)$ is the inverse of $x^{*}(s)$, i.e.\ the solution of $s^{*}(x^{*}(s))=s$. We stress the fact that the strict concavity of $J(s)$ implies the existence and uniqueness of the solution $s^{*}(x)$ of $J^\prime(s)=x$. Only in this case one has a one-to-one correspondence~(\ref{eq:slopes}) between the slopes of the cumulant GF $J(s)$ and the slopes of the rate function $\Psi(x)$.  

What is the meaning of these quantities? The answer is provided by~\eref{eq:EE_s}-\eref{eq:Jdiff} and the stationarity condition $\delta\EE_s[\rho^*_s]/\delta\rho=0$. Indeed,
\begin{equation}
x^{*}(s)=J^{\prime}(s)=\frac{\de}{\de s}\EE_{s}[\rho^{*}_{s}]=\int\underbrace{\frac{\delta\EE_s}{\delta \rho(\lambda)}[\rho^{*}_s]}_{=0}\frac{\partial\rho^{*}_s(\lambda)}{\partial s}\de \lambda+\frac{\partial\EE_s}{\partial s}[\rho^{*}_s]=F[\rho^{*}_{s}],
\label{eq:xstar}
\end{equation}
as anticipated in~\eref{eq:f(s)}.

The identity  \eref{eq:JPsi} can be informally written in the (almost) symmetric form
\be
J(s)-\Psi(x)=sx\ . \label{eq:sx}
\ee
This equation should be handled with care. In fact, in \eref{eq:sx}, there is only one independent variable: \emph{either} $s$ \emph{or} $x$. The relation between the conjugate variables $x$ and $s$ is ruled by~\eref{eq:slopes},
and the identity~\eref{eq:sx} is to be interpreted as either~\eref{eq:JPsi1} or as~\eref{eq:PsiJ1}.
In any case we can stipulate that
\barr
\cases{\phantom{+}\de J(s)=x^{*}(s)\de s , \qquad \text{$J(0)=0$},\\
-\de \Psi(x)=s^{*}(x)\de x , \qquad\! \!\text{$\Psi(x_0)=0$},} 
\label{eq:stip}
\earr
with $x_0=x^{*}(0)$. These equalities show that, when a large parameter ($N$ in our case) is involved, the Laplace and the Legendre transforms are intimately connected through the saddle-point approximation of the thermodynamic limit. 

The thermodynamic relations~\eref{eq:sx} and \eref{eq:stip} provide an alternative method to compute the large deviation functions.  The steps of this shortened method are
\begin{enumerate}
\item Solve the saddle-point equation~\eref{eq:optim} to get $\rho^{*}_s$;
\item Make the relation between $x$ and $s$ explicit by evaluating $x^{*}(s)=F[\rho^{*}_s]$ in~\eref{eq:f(s)}, and its inverse $s^{*}(x)$;
\item Compute the large deviation functions $J(s)$ and $\Psi(x)$ using the relations~\eref{eq:stip}. Hence we have
\be
J(s)=\int_0^s{x^{*}(s^{\prime})\de s^{\prime}},\qquad \Psi(x)=-\int_{x_0}^xs^*(x')\de x', \label{eq:new}
\ee
with $x_0=x^{*}(0)$.
\end{enumerate}
This procedure 
is usually many times faster than the standard route, as the following examples clearly demonstrate.
\paragraph{Example 1: Coulomb gas in a box.}  
As a first example we consider a Coulomb gas confined in the interval $[-1,1]$. The canonical measure of the $N$-particle gas at inverse temperature $\beta>0$ is
\be
\de P(\lambda_1,\dots,\lambda_N)=Z_{N,\beta}^{-1}\prod_{i<j}|\lambda_i-\lambda_j|^{\beta}\de\lambda_1\cdots \de\lambda_N,\qquad-1\leq\lambda_i\leq1.
\label{eq:jpdf_ex}
\ee
Let us take the center of mass of the gas $F=N^{-1}\sum_i\lambda_i$ as our natural observable. Clearly $-1\leq F\leq1$. 
We will derive the rate function $\Psi(x)$ of $F$ in the large $N$ asymptotics  $\mathcal{P}_N(x)\approx \exp(-\beta N^2 \Psi(x))$  using the shortened route. Note that $\Psi(x)=+\infty$ (zero probability) if $x<-1$ or $x>1$. 

The first step is to compute the saddle-point density $\rho_s^*(\lambda)$ of the effective energy functional~\eref{eq:functional_eff} with $f(\lambda)=\lambda$ and $V=0$. This optimization can be performed using, for instance, Tricomi's formula~\cite{Tricomi57}. Skipping details, one eventually finds that 
\be
\rho_s^*(\lambda)=\frac{(a+b-2\lambda)s+2}{2\pi\sqrt{(b-\lambda)(\lambda-a)}},\label{eq:saddle_ex}
\ee 
supported on the interval $[a,b]$ whose edges depend on $s$ as 
\begin{equation}\label{eq:edges_ex}
  [a,b]=
\cases{
  [1+2/s,1]  &{for $s\leq-1$},\\
 [-1,1]&{for $ |s|\leq1$},\\
   [-1,-1+2/s] &{for $ s\geq1$}.
}
\end{equation}
From~\eref{eq:saddle_ex}-\eref{eq:edges_ex} one easily computes $F[\rho^*_s]=\int f(\lambda)\de\rho^*_s(\lambda)$
\be\label{eq:Frho}
\int \lambda\de\rho^*_s(\lambda)=\frac{\pi}{\mathrm{i}}\mathop{\mathrm{Res}}_{z=\infty}\left\{z\frac{(a+b-2z)s+2}{2\pi\sqrt{(z-b)(z-a)}} \right\}=\frac{(a+b)}{2}-\frac{(a-b)^2}{8}  s.
\ee
By inserting the expressions of $a$ and $b$ from~\eref{eq:edges_ex} into~\eref{eq:Frho} one gets the explicit form of $x^*(s)=F[\rho^*_s]$  and its inverse $s^*(x)$:
\begin{equation}\label{eq:xs}
 x^*(s)=
\cases{
  1/(2s)+1  &{if  $s\leq-1$},\\
 -s/2  &{if  $|s|\leq 1$},\\
  1/(2s)-1 &{if  $s\geq1$},
}
\end{equation}
\begin{equation}\label{eq:s*x}
s^*(x)=\cases{1/[2(x-1)]  &{if $1/2\leq x\leq1$},\\
-2x  &{if  $-1/2\leq x\leq1/2$},\\
 1/[2(x+1)] &{if  $-1\leq x\leq-1/2$}.}
\end{equation}
According to formulae~\eref{eq:new}, a straighforward integration  of \eref{eq:xs}-\eref{eq:s*x} provides the large deviation functions 
\begin{equation}
\fl\qquad\quad\;   J(s)=\int_{0}^sx^*(s')\de s'=
  \cases{
  s+\log{\sqrt{-s}}+3/4   &{if $s\leq-1$},\\
 -s^2/4   &{if $-1\leq s\leq 1$},\\
  -s+\log{\sqrt{s}}+3/4   &{if $s\geq1$}.
  }
\end{equation}
\begin{equation}
\fl\qquad\quad   \Psi(x)=-\hspace{-2mm}\int\limits_{x^*(0)}^x \hspace{-1mm}s^*(x')\de x'=
  \cases{
  1/4+\log{\sqrt{2(1-x)}}   &{\hspace{-4mm}if  $1/2\leq x\leq1$},\\
 x^2  & {\hspace{-4mm}if $-1/2\leq x\leq1/2$},\\
  1/4+\log\sqrt{2(x+1)}   &{\hspace{-4mm}if $-1\leq x\leq-1/2$}.
  }
\end{equation}
Note that both the cumulant GF $J(s)$ and the rate function $\Psi(x)$ are \emph{not} analytic. More precisely, the third derivative of the free energy $J(s)$ (the rate function $\Psi(x)$) of the Coulomb gas is discontinuous at $s=\pm1$ (at $x^*(\pm1)=\mp1/2$). These non-analyticities correspond to phase transitions of the Coulomb gas. We will come back to this point in the last section.

\paragraph{Example 2: Planar approximation of quantum field theories.} Using Wick calculus it is possible to show that the cumulants (connected averages) of powers of traces of complex Gaussian matrices have a $1/N$ expansion in terms of maps with given genus. See~\cite{Zvonkin97}. To leading order in $N$, these cumulants are given by enumeration of planar diagrams (maps of genus zero). 

Let $M$ be a Gaussian matrix $\de\mu(M)\propto \exp(-\beta N\Tr M^2/4)$ and let $F(M)=N^{-1}\Tr M^4$. For $\beta=2$ (complex Gaussian model) the cumulants of $F(M)$ to leading order in $N$ enumerate the connected planar vacuum diagrams of the
$\varphi^4$-theory. In~\cite{Brezin78} the cumulant GF of these numbers (among other results) was computed using a Coulomb gas picture. Here we reproduce their result using the shortened version of the Coulomb gas method (beware of different notation: our Laplace parameter $s$ corresponds to $2g$ in the convention of~\cite{Brezin78}).

The Gaussian ensemble corresponds to a quadratic potential $V(\lambda)=\lambda^2/4$ in~\eref{eq:energy}. The spectral statistics we are interested in is $F=N^{-1}\sum f(\lambda_i)$ with $f(\lambda)=\lambda^4$. We compute the sum of connected vacuum diagrams $J(s)$  (the cumulant GF of the linear statistics $F$) using the Legendre duality. Notice that the Laplace transform $\widehat{\mathcal{P}}(s)=\int e^{-\beta N^2 s F(M)}\de\mu(M)$ is finite only for $s\geq0$. Hence, a priori, the first hypothesis of the G\"{a}rtner-Ellis theorem (Lemma~\ref{lem:GE}) is not satisfied. 

The saddle-point density $\rho^*_s$ (the minimizer of $\EE_s[\rho]$) can be computed explicitly using standard methods
\be
\rho^*_s(\lambda)=\frac{1}{\pi}\left[1/2+8sL(s)+4s\lambda^2\right]\sqrt{4L(s)-\lambda^2}
\ee
for $|\lambda|\leq 2\sqrt{L(s)}$ with $L(s)=(1/48s)(\sqrt{1+96s}-1)>0$. Using residue calculus as in the previous example one obtains
\be
x^*(s)=\int f(\lambda)\de\rho^*_s=L^2(s)[3-L(s)].
\ee
Insisting on the validity of the Legendre duality, an elementary integration gives:
\barr
J(s)&=\int_0^s x^*(s')\de s'
=\int_{L(0)}^{L(s)}L^2(3-L)\frac{(L-2)}{24 L^3}\de L\nonumber\\
&=\frac{1}{48}\Big(L(s)-1\Big)\Big(9-L(s)\Big)-\frac{1}{4}\log L(s) .
\earr
This cumulant GF was computed in~\cite[Eq.\ (20) and~(21)]{Brezin78} as difference of energy functionals $J(s)=\EE_s[\rho^*_s]-\EE_0[\rho^*_0]$. Our shortened route recovers this classical result with so little effort that we felt worth sharing.  Notice that $J(s)$ can be  analytically continued around $s=0$.  
The number of connected planar vacuum diagrams of the
$\varphi^4$-theory
are given by the derivatives at $s=0$ of the cumulant GF: $(-\beta)^{1-m}\partial_{ s}^{m}J(0)$, with $\beta=2$. The first few numbers are $2,36,1728,145152,17915904,\dots$. 

\subsection{Joint linear statistics and thermodynamic identities}
The generalization to $n>1$ linear statistics $\vec{F}(M)$ in~\eref{eq:nlinstat} 
is immediate. Now, $J(\vec{s})$ and $\Psi(\vec{x})$ are functions of $n$ real variables. Again we assume $J(\vec{s})$ strictly concave and differentiable so that the Legendre-Fenchel transform reduces to the simplest Legendre transform. Let $\rho^*_{\vec{s}}$ be the minimizer of the mean-field energy functional~\eref{eq:functional_eff_multi}. The multidimensional analogue of~\eref{eq:sx} is
\be
J(\vec{s})-\Psi(\vec{x})=\vec{s}\cdot\vec{x},
\ee
where $\vec{x}=\vec{x}^*(\vec{s})$ has components 
\be
x_i^*(\vec{s})=F_i[\rho^*_{\vec{s}}]=\int f_i(\lambda)\de\rho^*_{\vec{s}}(\lambda) \qquad (i=1,\dots,n), \label{eq:sx_multi}
\ee  
and $\vec{s}=\vec{s}^*(\vec{x})$ is its inverse map.
We can write the differential equations 
\barr
\de J(\vec{s})=\sum_{i=1}^nx_i^*(\vec{s})\de s_i, \qquad
\de \Psi(\vec{x})=-\sum_{i=1}^ns_i^*(\vec{x})\de x_i,
\label{eq:JPsimulti}
\earr
supplemented by the `initial conditions' 
\be
\quad J(\vec{0})= \Psi(\vec{x}_0)=0\qquad\text{ where  }\quad\vec{x}_0= \vec{x}^*(\vec{0}).
\ee Hence the computation of $J(\vec{s})$ and $\Psi(\vec{x})$ amounts to finding the saddle-point density $\rho^{*}_{\vec{s}}$, computing $F_i[\rho^*_{\vec{s}}]$ ($i=1,\dots,n$) in~\eref{eq:sx_multi} and then integrating the differential forms~\eref{eq:JPsimulti}.

The identities~\eref{eq:JPsimulti} have been employed \cite{Cunden14,Grabsch15,Cunden15} and explicitly stated \cite{GrabschPV,Grabsch15,Cunden15} in a few previous works on joint linear statistics on matrix models. However, their precise LDT conditions of applicability, as well as the implications for weak phase transitions and evaporation phenomena (discussed below in Section \ref{sec:discussion}) do not seem to have been examined elsewhere.

\subsection{Back to thermodynamics}

In order to convey the main ideas of this method and show its relation with thermodynamics, we discuss the $n=2$ case, i.e.\ the case of two linear statistics $F_1(M)$ and $F_2(M)$. Now $\vec{x}=(x_1,x_2)$ and $\vec{s}=(s_1,s_2)$. Once the variational problem $\delta\EE_{s_1,s_2}[\rho^*_{s_1,s_2}]/\delta\rho=0$ is solved, one gets 
\begin{equation}
x_1^*(s_1,s_2)=F_1[\rho^*_{s_1,s_2}], \qquad x_2^*(s_1,s_2)=F_2[\rho^*_{s_1,s_2}],
\end{equation} 
and the pair of inverse maps $s_1^*(x_1,x_2)$, $s_2^*(x_1,x_2)$. The differential relations~\eref{eq:JPsimulti} in this case read
\begin{equation}
\cases{\phantom{+}\de J(s_1,s_2)=x_1^*(s_1,s_2)\de s_1+x_2^*(s_1,s_2)\de s_2,\\
-\de\Psi(x_1,x_2)=s_1^*(x_1,x_2)\de x_1+s_2^*(x_1,x_2)\de x_2,}
\label{eq:JPb}
\end{equation}
or, for short,
\begin{equation}\label{eq:thermoback}
J-\Psi = x_1 s_1 + x_2 s_2 .
\end{equation}
The reader should have recognized the familiar Maxwell relations for thermodynamic potentials~\cite{Huang}. This should not come as a surprise, as LDT may be really seen as the proper mathematical framework in which statistical mechanics can be formulated rigorously (several authors have elaborated on this idea; see e.g.~\cite{Ellis99}).  Let us then elaborate briefly on this point. In thermodynamics, temperature $T$ and pressure $p$ say, are the control parameters of (the Lagrange multipliers associated with) entropy $S$ and volume $V$, respectively. A classical $(p,V,T)$ system is conveniently described in terms of the internal energy $U(S,V)$ or the Gibbs free energy $G(T,p)$. Bearing in mind the different sign convention in the thermodynamic tradition, the analogues of~\eref{eq:JPb} are~\cite{Huang}:
 \begin{equation}
\cases{\phantom{+}\de U(S,V)=T\de S-p\de V,\\
-\de G(T,p)=S\de T-V\de p,}
\label{eq:JPthermo}
\end{equation}
 or 
\begin{equation}
U-G=TS-pV,
\end{equation}
for short (cf.~\eref{eq:thermoback}), where
\be
\cases{T=\left(\frac{\partial U}{\partial S}\right)_V, \quad p=-\left(\frac{\partial U}{\partial V}\right)_S,\\
V=\left(\frac{\partial G}{\partial p}\right)_T, \quad S=-\left(\frac{\partial G}{\partial T}\right)_p.}
\ee

\section{Discussion and Outlook}
\label{sec:discussion}
In the previous sections we have presented a revision of the Coulomb gas method for linear spectral statistics on random matrices. Thanks to the contraction principle (Lemma~\ref{lem:contraction}) it is possible to retrieve information from the behavior of the `gas of eigenvalues' and establish a LDP for linear statistics. In order to compute the large deviation functions through  the G\"artner-Ellis theorem (Lemma~\ref{lem:GE}), we have made an extensive use of the Legendre duality.

The bottleneck of the standard method summarized in Section~\ref{sec:method} is the evaluation of the energy functional~\eref{eq:functional_eff_multi} at the saddle-point density, which entails the computation of many integrals. Here we have illustrated a computationally
simpler approach. This shortened method only requires the explicit relation between the real variables $\vec{x}$ and the Laplace variables $\vec{s}$ from~\eref{eq:sx_multi} and its inverse. The workload can be drastically reduced by exploiting the differential relations~\eref{eq:JPsimulti}, with the welcome consequence that the otherwise daunting task of finding \emph{joint} large deviations is also made possible (see e.g.\ the recent works~\cite{Cunden14,Cunden15,Grabsch15}). The method can also be applied almost verbatim to real spectral statistics on invariant normal non-Hermitian ensembles (see~\cite{Cunden15b} for an application on complex Ginibre matrices~\cite{Ginibre}).

As already mentioned, the rate function $\Psi(\vec{x})$ of a spectral linear statistics provides an overall description of the fluctuations about the most probable values. If $J(\vec{s})$ is \emph{analytic} at $\vec{s}=0$, the mixed \emph{cumulant} of  $F_1(M),\dots,F_n(M)$ to leading order in $N$ are given by  
\be
\fl\qquad\quad \<F_1(M)^{m_1}\cdots F_n(M)^{m_n}\>_c=\left(-\frac{1}{\beta N^2}\right)^{m_1+\cdots+m_n-1}\!\!\partial_{ s_1}^{m_1}\cdots \partial_{s_n}^{m_n}J(\vec{s})\Bigl|_{\vec{s}=\vec{0}}.
\label{eq:mixed_cumulants}
\ee
(We used this formula in Example~2.)  From~\eref{eq:mixed_cumulants} one sees that the averages are $\<{F_i(M)}\>=\Or(N^0)$, the covariances are $\mathrm{Cov}(F_i(M),F_j(M))=\Or(N^{-2}$) and the higher order cumulants ($\sum_im_i>2$) decay faster with $N$. It is often stated that a quadratic minimum of $\Psi(\vec{x})$ implies a Central Limit Theorem for the fluctuations of $F_1(M),\dots, F_n(M)$ around their typical values. In fact, this is not true in general and stronger conditions are required to infer a Central Limit Theorem~\cite{Bryc93}. 

In what follows, we first briefly discuss the regularity properties of the cumulant GF and the associated phase transitions within the underlying Coulomb gas framework. Then we conclude with a discussion of some subtleties of the Coulomb gas method.

\subsection{Phase transitions in 2D Coulomb gases}
The regularity properties of $\Psi(\vec{x})$ and $J(\vec{s})$ are ultimately determined by the smoothness of the map $\R^n\ni\vec{s}\longmapsto\rho^*_{\vec{s}}\in\mathcal{M}(\R)$. 
Coming back to a single statistics ($n=1$) for simplicity, it has been observed in several problems that the cumulant GF $J(s)$ is \emph{not} real analytic (see Example~1). Since $J(s)$ is the excess free energy~\eref{eq:Jdiff} of the associated 2D Coulomb gas, the non-analyticity points $s_{\mathrm{cr}}$ correspond to \emph{phase transitions} of the Coulomb gas system. The \emph{order} of the phase transition at $s_{\mathrm{cr}}$ (in the sense of Ehrenfest) is the smallest integer $\ell\in\mathbb{N}$ such that $\partial_s^{\ell}J(s)$ is discontinuous at $s_{\mathrm{cr}}$. 

Phase transitions in matrix models abound. These transitions are remarkably weak, usually of third-order (the third derivative of $J(s)$ being discontinuous at some $s_{\mathrm{cr}}$; in the Example~1 we have seen such a discontinuity of $J(s)$ at $s_{\mathrm{cr}}=\pm1$). See~\cite{Colomo13,DePasquale10,Douglas93,Facchi08,Facchi13,Gross80,Nadal10,Vivo08}. Similar third-order phase transitions, driven by the very same mechanism (known as `hard-wall mechanism', see the broad survey~\cite{Majumdar14}) have been also observed for joint ($n>1$) linear statistics, revealing rich phase diagrams~\cite{Cunden15}. Third-order phase transitions associated to the merging/splitting of the gas density in multiple supports have also been studied~\cite{Vivo08}. Recently a fourth-order transition (driven by a change of topology mechanism) has been detected in the Ginibre ensemble~\cite{Cunden15b}. 

For a single statistics, the identity $J^{\prime}(s)=x^*(s)$ shows that the Coulomb gas undergoes an $\ell$-th order phase transition at $s_{\mathrm{cr}}$ if $\partial_s^{\ell-1}x^{*}(s)$ is discontinuous at $s_{\mathrm{cr}}$ where $x^{*}(s)=\int f(\lambda) \de \rho^*_s(\lambda) $. For instance, in our Example 1, $x^*(s)$ in Eq.~\eref{eq:xs} has a continuous first derivative; its second derivative  $\partial_s^{2}x^{*}(s)$ is discontinuous at $s_{\mathrm{cr}}=\pm1$ and hence the non-analyticities of the free energy $J(s)$ are of third-order. Also note that, since $-\Psi'(x)=s^*(x)$, the rate function $\Psi(x)$ and the cumulant GF $J(s)$ have the same regularity. 

The study of phase transitions in invariant matrix models is very much a topic of current research. A natural question is to know whether the free energy $J(s)$ associated to a linear statistics on a invariant matrix models has critical points and to characterize the order of the transitions. Physically one certainly expects a relation between the particular behavior of the gas density $\rho_s^*$ and the arising non-analyticities of the corresponding thermodynamic observables  (see Example 1). However, this relation is not completely understood in general, and each particular case requires explicit working. 
Interestingly,
the identities discussed in this article can be used to get new insights on the emergence and characterization of these weak phase transitions~\cite{CFLV}. 

\subsection{Breakdown of the method: discontinuous statistics and evaporation phenomena}
At this point it is worth mentioning the limitations of the method based on Legendre duality.  The procedure outlined in Section~\ref{sec:shortcut} relies on the Contraction Principle (Lemma~\ref{lem:contraction}) and  the G\"artner-Ellis theorem (Lemma~\ref{lem:GE}).

The method does not work in general for \emph{discontinuous} statistics -- a paradigmatic example is the fraction of eigenvalues $F(M)=N^{-1}\sum_i\chi_I(\lambda_i)$ in a fixed interval $I$. This is a fundamental obstruction because in these cases it is not even guaranteed that a LDP exists: the contraction principle does not work in general for $F(M)=N^{-1}\Tr f(M)$ when $f$ is not continuous. In many cases~\cite{Dyson62,Majumdar11,Marino14}, the hallmark of this obstruction is a logarithmic contribution to the variance $\mathrm{Var}( F(M))\sim (\log N)/N^2$, in contrast to the $\Or(N^{-2})$ general behavior for smooth linear statistics discussed above.

More interesting limitations of the method are related to the properties of the cumulant GF $J(s)$. The G\"artner-Ellis theorem (Lemma~\ref{lem:GE}) is not applicable if the so-called \emph{steepness condition} $\|\vec{\nabla}J(\vec{s})\|\to\infty$ (as $s\to\partial\mathcal{D}$) is not verified. 
In this case, only a `local' version of the G\"artner-Ellis theorem holds since the map $\vec{s}^*(\vec{x})$ is not globally defined. This is the case for many if not all previous works on linear statistics~\cite{Cunden14,DePasquale10,Nadal10,Facchi13,Texier13,Grabsch15} where a phenomenon of evaporation of eigenvalues in random matrices was claimed.

To simplify the discussion  consider a single statistics $F(M)=N^{-1}\Tr f(M)$ ($n=1$) and  suppose again that $J(s)$ is strictly concave (so that $J'(s)=x^*(s)$ is a strictly decreasing  function). For concreteness suppose that $J(s)$ is defined for $s\geq s_0$ and that $J'(s_0^+)=C$ (failure of the steepness condition). Hence $J'(s)=x^*(s)\leq C$. The inverse function $s^*(x)=-\Psi'(x)$ is therefore defined only for $x\leq C$ and the rate function $\Psi(x)$ cannot be recovered by integration~\eref{eq:new} when $x$ is larger than $C$. This means that the rate function computed as Legendre transform of the non-steep cumulant GF provides only a partial description (for values $x\leq C$) on the limiting distribution $\mathcal{P}_N(x)\approx \exp(-\beta N^2 \Psi(x))$ of $F(M)$.

Note that $F(M)=N^{-1}\sum_i f(\lambda_i)$ is a sum of strongly correlated random variables. Several papers in the
physics literature ran into a similar obstruction: the constrained Coulomb gas calculation provides the rate function $\Psi(x)$ of a LDP with speed $\Or(N^2)$ as long as $x$ is less (say) than a critical value $C$. In this regime, the typical eigenvalue distribution constrained by $F[\rho]=x\leq C$ is the solution of a variational problem like~\eref{eq:optim}, and all the eigenvalues `democratically' contribute to satisfy the constraint. Guided by numerics and physical arguments, in several problems it has been argued that the statistics of values of $F(M)$  larger than the critical value $C$ is instead driven by evaporation phenomena: in the large $N$ limit, a single eigenvalue splits off from the continuous density of $M$. In this scenario, large values of the sum $F(M)$ are dominated by the contribution of a single summand. The hallmark of this is a change of speed in the large deviation estimate, usually from $\Or(N^2)$ in the `democratic regime' to $\Or(N)$ in the regime where a single eigenvalue carries a macroscopic contribution to $F(M)$. 

A thorough review of these works, in the modern viewpoint adopted in this paper, reveals that the aforementioned obstruction can be essentially traced back to the non-steepness of the cumulant GF of $F(M)$.  Therefore, {a failure of the steepness condition may correspond to a `change of speed' in the LDP driven by evaporation phenomena}. This phenomenon is not restricted to random matrices and is naturally explained 
by a careful consideration of LDT detailed prescriptions.

\ack
FDC acknowledges  support  from EPSRC Grant No.\ EP/L010305/1. PF acknowledges  support  from PRIN 2010LLKJBX on ``Collective quantum phenomena: from strongly correlated systems to quantum simulators''.  PV acknowledges  support  from  EPSRC  Centre  for  Doctoral  Training  in Cross-Disciplinary Approaches to Non-Equilibrium Systems (CANES), and gratefully acknowledges several discussions and exchanges with C. Texier and A. Grabsch on their results. FDC and PF  received furthermore partial support from the Italian National Group of Mathematical Physics (GNFM-INdAM).

\vspace{1cm}
%%%%%%%%%%%%%%%

\end{document}